\documentclass[12pt]{spieman}  

\usepackage{amsmath,amsfonts,amssymb}
\usepackage{graphicx}
\usepackage{setspace}
\usepackage{tocloft}
\usepackage{lineno}
\usepackage{orcidlink} 
\usepackage{bookmark}

\newcommand\tool{\texttt{corosims}}
\newcommand\cgisim{\texttt{cgisim}}
\newcommand{\lamOverD}{\ensuremath{\lambda/D}}

\newcommand{\taucet}{$\tau$ Ceti}
\newcommand{\degree}{\ensuremath{^\circ}}
\newcommand{\vmagasec}{\textit{V}~mag/arcsec$^2$}

\newcommand{\tentoe}{\ensuremath{10^{-8}}}
\newcommand{\tenton}{\ensuremath{10^{-9}}}

\title{Roman coronagraph simulations of exozodi observations in the presence of wavefront errors }

\author[a]{Jorge Llop-Sayson\orcidlink{0000-0002-3414-784X}}
\author[a]{Vanessa P. Bailey\orcidlink{0000-0002-5407-2806}}
\author[b]{Justin Hom}
\author[a]{John Krist}
\author[a]{Bertrand Mennesson}
\author[c]{Samantha N. Hasler}
\author[d]{Alexandra Z. Greenbaum\orcidlink{0000-0002-7162-8036}}
\author[a]{A~J Eldorado Riggs}
\author[a]{Geoffrey Bryden\orcidlink{0000-0001-5966-837X}}
\affil[a]{Jet Propulsion Laboratory, California Institute of Technology, 4800 Oak Grove Dr., Pasadena, CA 91109, USA}
\affil[b]{Steward Observatory and Department of Astronomy, University of Arizona, 933 N Cherry Ave., Tucson AZ 85721, USA}
\affil[c]{Massachusetts Institute of Technology, 77 Massachusetts Ave., Cambridge, MA 02139, USA}
\affil[d]{IPAC, Caltech, 1200 E. California Blvd., Pasadena, CA 91125, USA}

\cftpagenumbersoff{figure}
\cftpagenumbersoff{table} 

\begin{document}
\maketitle


\begin{abstract}

The Coronagraph Instrument on board of the Nancy Grace Roman Space Telescope will demonstrate key technologies that will prepare the ground for the Habitable Worlds Observatory. The current predictions for the Roman Coronagraph's detection limit range from $\sim10^{-8}$ to a few \tenton, which would allow for groundbreaking science, such as potentially imaging Jupiter-like exoplanets around mature stars or imaging exozodi disks in scattered light. However, the performance of the instrument depends on many factors. Simulating images with varying optical error sources can help us connect instrument and observatory performance to potential science yield. 
Here we present a new tool, \tool, to simulate observations of arbitrary astrophysical scenes with the Roman Coronagraph in the presence of evolving wavefront errors. This tool wraps around the Roman Coronagraph PROPER diffraction model and detector simulator.
We use \tool~to investigate the potential degeneracy between jitter-induced speckles and both hot and warm exozodi disk structures.  First, we simulate observations of warm exozodi around \taucet, with varying jitter levels. 
We predict that with nominal post-correction pointing jitter performance ($\sim$0.3 mas RMS), the Roman Coronagraph should be sensitive to 12$\times$ zodis worth of dust, assuming a face-on (worst case scenario) inclination. We further predict that its sensitivity degrades to 35$\times$ zodis if jitter on-target is 3$\times$ worse than the nominal value. This estimate assumes the best-modeled wavefront control and stability values from the Roman project, {including} additional ``model uncertainty factors'' on performance. We find that, while jitter hinders warm exozodi detection by increasing the noise, jitter residuals are unlikely to result in a false positive, particularly if two-color imaging is conducted. However, if a faint, hot exozodi disk falls $<\lambda/D$ projected separation, it may not be distinguishable from jitter-induced speckle residuals of comparable brightness. Finally, we discuss the degeneracies induced between 
flux and separation retrieved near the inner working angle due the sharp edge of the Roman Coronagraph's focal plane mask. 

\end{abstract}


\begin{spacing}{2}   

\section{Introduction} \label{sec:intro}
The Nancy Grace Roman Space Telescope, set to launch no later than May of 2027, will be equipped with a Coronagraph Instrument (Roman Coronagraph hereafter), and will be the first space coronagraph with active wavefront sensing and control. As such it is expected to reach unprecedented levels of starlight suppression. This technology demonstration is a critical step for the maturation of the Habitable Worlds Observatory (HWO), NASA's mission concept under development in response to the recommendations from the Decadal Survey in Astronomy and Astrophysics \cite{Decadal}. To achieve its monumental task of detecting biosignatures in the atmospheres of Earth-like exoplanets, HWO will need to bridge a myriad of technology gaps. The Roman Coronagraph, as a technology demonstrator \cite{Mennesson2022}, is set to mature many of these. Some of the technologies that will be tested for the first time in a space coronagraph are: large-format deformable mirrors (DMs) \cite{Cavaco2014}, photon counting electron-multiplying CCDs (EMCCDs) \cite{Morrissey2023,Nemati2020,Harding2016}, high performance coronagraph masks and optics \cite{Trauger2016,Carlotti2013,Riggs2021,Gersh-Range2022,Bala2015}, and wavefront sensing and control systems \cite{Shi2016}. 
During the first 18 months of operations, the Roman Coronagraph will have 90 days of operational time. The Threshold Requirement is to reach a detection limit of $10^{-7}$ {flux ratio after post-processing} at 6~\lamOverD, but current performance predictions anticipate it could reach between \tentoe~and a few \tenton~detection limit at 4~\lamOverD\cite{Bailey2023}. %

The remarkable complexity of the Roman Coronagraph requires precise modeling of the system in order to predict the final performance of the instrument. 
To address this, many efforts, in parallel to the development of the instrument, focused on detailed simulations of the instrument+observatory system performance  
\cite{Krist2015,Douglas2020,Zhou2016,Zhou2020}. Ref.~\citenum{Krist2023} gives a summary of Roman Coronagraph modeling. In their work, the end-to-end modeling approach for the project is presented: the simulations are divided in a structural, thermal and optical performance model (STOP), and the coronagraph instrument diffraction model. The STOP model contains the structural and thermal finite elements model of the observatory and instrument, and the ray-trace optical model \cite{Saini2017}; it computes the disturbances for a given observation scenario, e.g. reference acquisition and science telescope roll observations. Running this model is computationally very expensive and in practice rarely done. The STOP model predicts the wavefront aberrations and optical element shifts that are fed to the diffraction model. With this set of errors from the STOP model, the diffraction model of the coronagraph computes the final images for a given observation scenario. The diffraction model of the Roman Coronagraph is implemented with the PROPER library \cite{Krist2007} and was made publicly available in a user-friendly wrapper, \cgisim,~
which allows for the simulation of Roman Coronagraph images with the right detector‐sized pixels and includes the wavelength dependent throughput of the instrument. 

The most current pre-flight performance estimates predict an achievable {detection limit} in the order of few times \tenton. This would allow, for the first time, the imaging of exoplanets analogous to Jupiter in reflected light \cite{Lacy2019, Batalha2018}. This {detection limit} would also allow resolving exozodi disks in reflection in the visible for the first time \cite{Schneider2016,Douglas2022}. Exozodi science with the Roman Coronagraph is a particularly interesting science case in the context of precursor science for HWO\cite{Mennesson2019a,Mennesson2019b}. Indeed, exozodiacal light is an astrophysical source of noise that jeopardizes the characterization of exo-Earth atmospheres \cite{Roberge2012,Stark2015}; the dust signal can obscure the planet or hinder noise estimation. In certain cases, when dust is trapped in clumps and mean motion resonance structures, just like our own zodiacal cloud \cite{Reach1995}, these can mimic the signal of a point source \cite{Defrere2012,Currie2023}. As stated by NASA's Exoplanet Exploration Program (ExEP) in their Science Gap List (SCI-11),~
``reducing the uncertainty in the median and distribution of exozodi levels and addressing the risk that the presence of hot dust may affect scattered light levels in the HZ'' is a key precursor science case for HWO. The Roman Coronagraph will be uniquely equipped to address this precursor science.

Furthermore a warm disk's structure, morphology, chemical composition and particle size distribution all attest to the rich history of the evolution of a planetary system. In our own Solar system, the presence of the zodiacal cloud is due to dust replenishment from dust-rich asteroid collisions and comets \cite{Nesvorny2011}. At an earlier age, during the late heavy bombardment, the composition of the warm cloud would have been abundant in water and other volatiles \cite{Gomes2005}, which would have fed the Earth with the necessary material to bring about life on its surface. The architecture of a disk also hints at the presence of larger bodies, and could help in the search of planets in the habitable zone \cite{Stark2008}. 

In this work we present a tool to produce realistic simulations of astrophysical scenes, and we apply it to the observation of exozodis. In Sec.~\ref{sec:tool_description} we present this tool, \tool, in Sec.~\ref{sec:tauceti_disk_simus} we apply \tool~to the case of \taucet~in order to predict sensitivity limits to different cases of telescope pointing jitter performance and disk inclinations. In Sec.~\ref{sec:toymodels_degeneracy}, we discuss some possible sources of confusion in the context of exozodi detection and characterization for the Roman Coronagraph.

\section{Simulation Tool} \label{sec:tool_description}
In this work we present \tool, a new simulation tool for the Roman Coronagraph. 
\tool~enables simulating astrophysical scenes with arbitrary amounts of residual post-correction pointing jitter (hereafter \textit{jitter}) and other wavefront errors for the different instrument modes of the Roman Coronagraph. This tool is conceived to allow simulations ranging from single dark hole images of the instrument to the simulation of full astrophysical scenes, containing speckle series for a set of scenes and including detector noise.
\tool~wraps around \cgisim~\cite{Krist2023}, the state-of-the-art propagation tool for the instrument, which contains the latest Roman Coronagraph optical propagation model, and the databases of optics throughputs, filters, masks, etc. One of the main motivations for the development of \tool~is to consolidate a package where the simulation of astrophysical scenes with arbitrary levels of jitter and other instrumental errors are unified and made user friendly. 

{The three main features of \tool~with respect to using \cgisim~are}:
\begin{itemize}
    \item The implementation of jitter and drift. 
    \item A convenient way of simulating an astrophysical scene through the instrument; with addition of planets and disks, in addition to using spectral models from \texttt{pysynphot} or custom spectra. A function in the code convolves an arbitrary scene with the instrument field dependent point response function (PRF).
    \item A user-friendly way of defining {astrophysical} scenes and simulating speckle series with evolving wavefront errors and optical drifts, with the future goal of implementing wavefront sensing and control (WFSC) with FALCO\cite{riggs2018}.
\end{itemize}

\subsection{Code Structure}\label{sec:code_structure}
In a similar philosophy to \texttt{pynrc}~\cite{Leisenring_pyNRC_Python_ETC_2024}, the code in \tool~is largely divided in two classes: (1) the \textit{core} class, which contains all functions and variables related to the simulation of the instrument, and the \textit{observations} class, which contains information about the scene and observation details. 

The core class includes the calls to \cgisim, which generally consist of the computation of the electric field at the image plane of the Roman Coronagraph. The electric field cube, consisting of the different wavelengths propagated through the optical model, are then rearranged and scaled according to the source specification, which are contained in a class attribute that manages the source details. Jitter is added, if specified by the user, before collapsing the cube and computing the intensity. The detector noise is added with \texttt{emccd\_detect}
, the EMCCD simulation package. The current version of the code does not allow for the computation of polarization aberrations, i.e. aberrations caused by the differential aberrations seen by orthogonal polarization states. This is due to \cgisim's output structure, which would have to be modified in order to allow different polarization states in the electric field mode. 


The observations class handles the scene and observation scenarios. The user can define a set of scenes with point sources or user-provided objects that are put through the simulator to obtain an image or a series of images. The user can define a set of wavefront errors for different timesteps. The observations class has a built-in convolution function that performs the PRF convolution operation of an arbitrary image with the Roman Coronagraph PRF of the corresponding mode and bandpass. The grid of PSFs needed to perform the convolution must be computed with the \texttt{compute\_jitter\_EFs} function, or can be downloaded from the repository's data directory: \url{https://doi.org/10.5281/zenodo.15586669}{. This grid of PSFs is different from the one needed to simulate the jitter, introduced in the following section.}

 
\subsection{Jitter and Drift}\label{sec:jitter}
\begin{figure}[t!]
   \begin{center}
   \begin{tabular}{c} 
   \includegraphics[height=10.0cm,trim={5cm 5cm 5cm 5cm}]{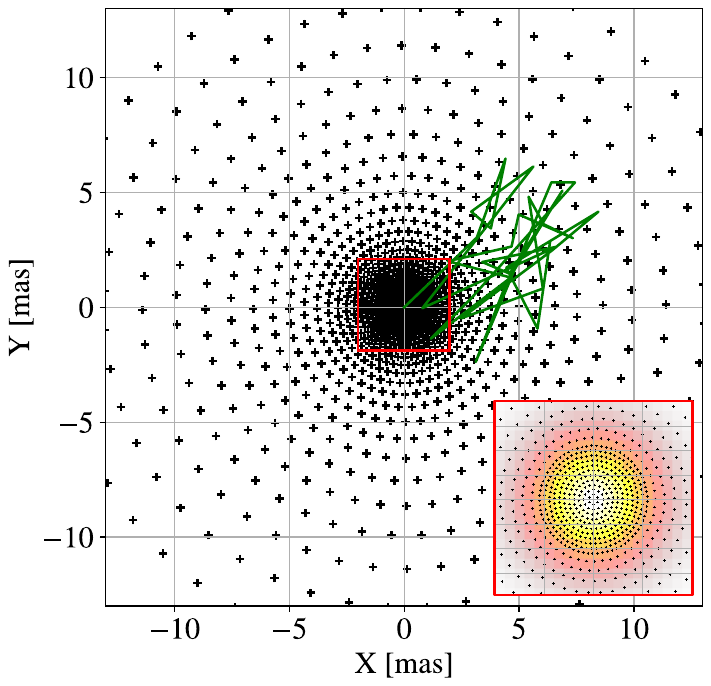}
   \end{tabular}
   \end{center}
   \caption{
   \label{fig:jitter_cloud}
   The crosses indicate the positions where the $\delta$EF are computed to simulate jitter and drift as explained in Sec.~\ref{sec:jitter}. Finer resolution is required near the center of focal pane mask, where the variability of the PSF structure is more important and where the object is expected to {spend} most of the acquisition time. This representation is for the case of the HLC mode. The cloud centered at [0,0] indicates a representative pre-LOWFSC jitter for the first 32 hours of the Observing Scenario 11 (OS11)\cite{Krist2023}. The green line represents the pre-LOWFS drift of the instrument for the first 10 hours of OS11. {The inset shows a zoomed in view of the inner region where a finer resolution of $\delta$EF is computed.}} 
\end{figure}

Line-of-sight errors can be categorized in jitter or drift: the former is fast (typically $>$1 HZ) tip-and-tilt motions with zero mean, while the latter is a slower motion with non-zero mean. Jitter and drift require special treatment in order to produce reliable simulations. This is due to the fact that the {point spread function (PSF)} structure varies significantly behind the FPM. The treatment of jitter is thoroughly explained in Section 5.7 of Ref.~\citenum{Krist2023}. In short, a precomputed cloud of $\delta$EFs (electric field changes with respect to an unaberrated EF$_0$) in the vicinity of the FPM center is used as a weighting function for a jitter instantiation defined as a 2D Gaussian function. Since faint objects are expected to yield larger jitter than the nominal case of bright stars, we, thus, define a larger cloud of $\delta$EF which allows for larger jitter and drift. Figure~\ref{fig:jitter_cloud} shows the distribution of points around the center of the FPM where the $\delta$EFs are computed in the case of the Hybrid Lyot Coronagraph (HLC)\cite{Seo2016} band 1 mode.

The treatment of drift is done in an analogous way; instead of a Gaussian function, the drift is represented by a line function. Since drift and jitter happen simultaneously, the  drift function is convolved with the jitter function. In Figure~\ref{fig:jitter_cloud} we show a representative pre-LOWFS correction jitter amplitude and a drift \textit{walk} from OS11 data. For a given time-step, the line from the drift is convolved with the jitter weight function to give a representative $\delta$EF for simulating jitter combined with drift. {The PSF grid shown in the figure is sufficiently large to accommodate significant drift.} {The expected jitter for the Roman Coronagraph is under 1 mas for \textit{V}~mag $<$5 targets, for which the drift is expected to be negligible with respect to the jitter amplitude \cite{Dube2022}.} 

The performance of the LOWFSC, and thus the amount of residual jitter and of wavefront error modes Z4-11 (Zernike noll index), depends on the amount of photons available at the LOWFS camera in order to sense the wavefront \cite{Shi2016, Shi2018, Dube2022}. Therefore, there exists a dependency of the amount of these low-order wavefront error residuals with the brightness of the target. Testbed experiments\cite{Shi2018} explored the magnitude dependence of Z2-Z4 sensing performance for guide stars $M_V \lesssim 5$, the required operating range for Roman Coronagraph. A future release of \tool~may add a simplified model of the LOWFS residuals as function of stellar magnitude for different coronagraph modes.

\subsection{Validating against OS11}
In order to validate \tool~we compare its outputs against the benchmark simulation time series, the Observing Scenario 11 (OS11)\cite{Krist2023}. {Both OS11 and \tool~utilize the proper model to perform the optical propagation through the system; however, the implementation of the jitter and the DM~shapes is different. This validation aims at confirming that the evolving jitter inputs result in similar changes in both OS11 and \tool, and that the normalized intensities are similar. Although there is no real data from the instrument in a relevant contrast regime, past efforts have validated similar simulation tools with testbed results\cite{Zhou2019}.} 
The OS11 package includes a timeseries of speckle images, or speckle series, of the observing scenario with its corresponding input series of wavefront errors. This makes it possible to reproduce the speckle series with the same inputs.  However, the DM~shapes used to obtain these speckle series {were} not provided {due to export control}, so it is not possible to produce the exact same speckle structure. Because the latest version of \cgisim, version 4.0.1, was released after the release of OS11, and it includes major changes in the optical model, we use the previous version of \cgisim, version 3.1, to make a more relevant comparison of our code to the results from OS11. {The rest of the paper uses version 4.0.1.}

\begin{figure}[t!]
   \begin{center}
   \begin{tabular}{c} 
   \includegraphics[height=7.0cm]{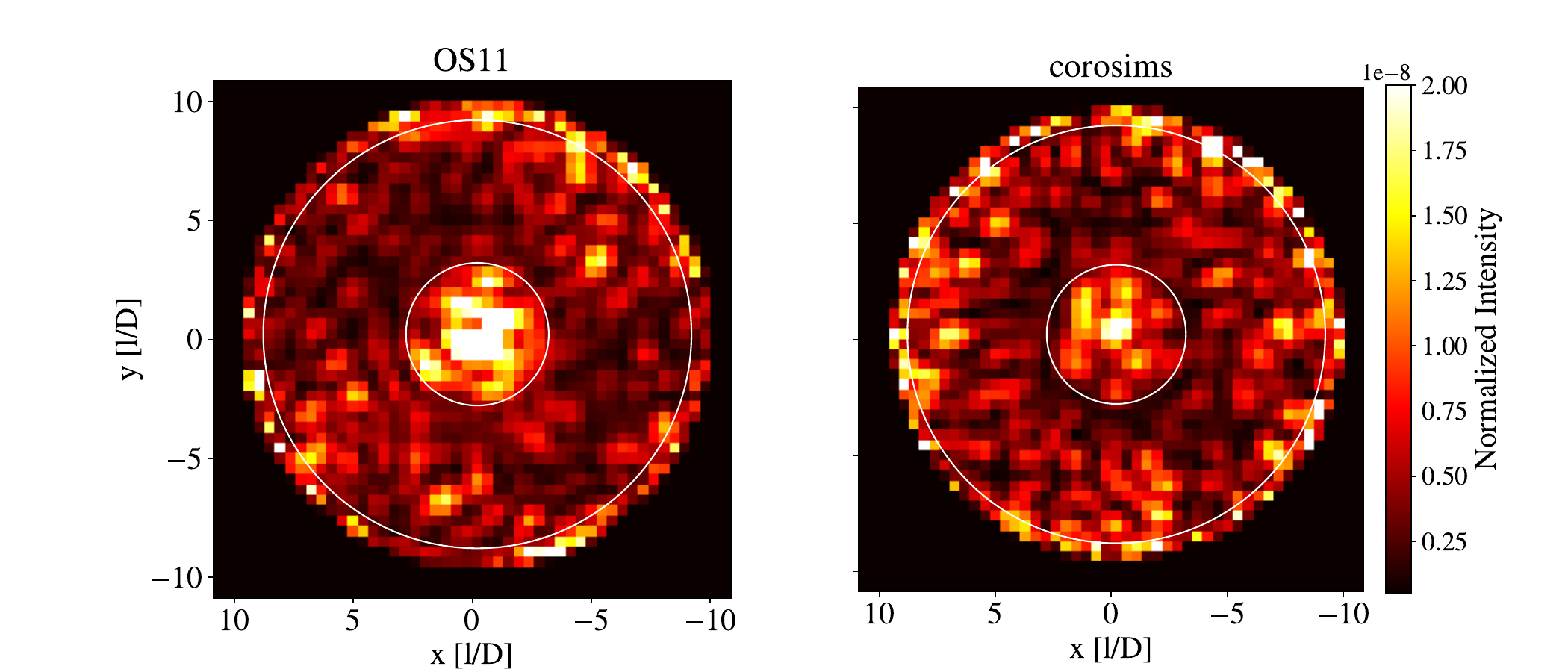}
   \end{tabular}
   \end{center}
   \caption{
   \label{fig:os11dh} 
   Dark hole images from the OS11 data package batch 0 (\textit{left}), and from \tool (\textit{right}), generated with the same inputs{, except for the jitter implementation and deformable mirror shapes}. The differences between speckles structures is due to the use of different deformable mirror shapes. White lines indicate inner and outer working angles (IWA and OWA), at 3 and 9~\lamOverD, respectively.} 
\end{figure}

\begin{figure}[t!]
   \begin{center}
   \begin{tabular}{c} 
   \includegraphics[height=8.0cm]{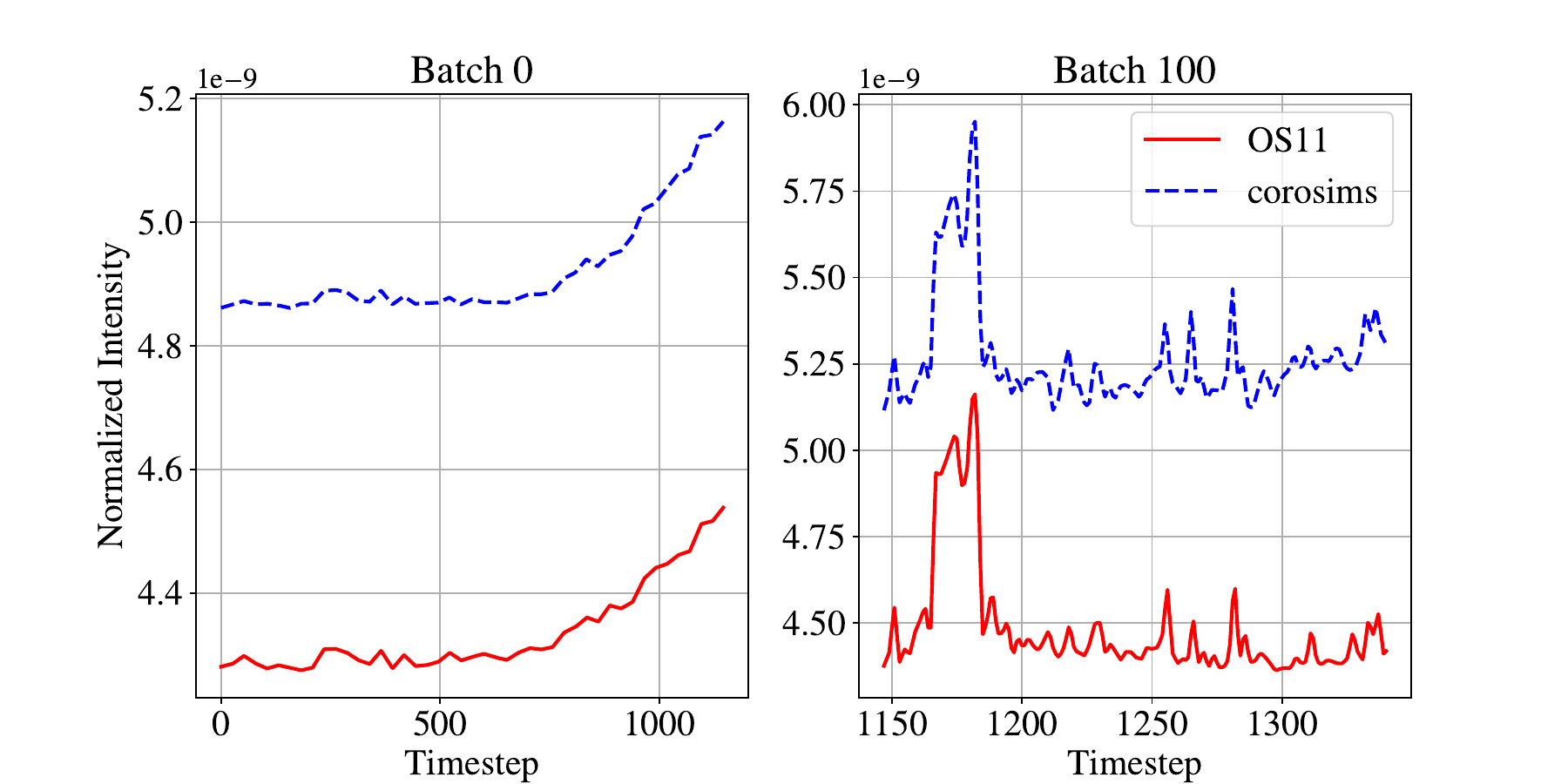}
   \end{tabular}
   \end{center}
   \caption{
   \label{fig:os11ni} 
   Normalized intensity comparison between the simulation package OS11 and the results from simulations with \tool~for the same inputs. Two different batches are shown: batch 0 (\textit{left}), from images of the first reference acquisition, batch 100 (\textit{right}), from images of the first roll science acquisition. The small difference in flux is due to the different DM solutions used,  which result in a different normalized intensity. The uptick in flux in the dark hole at the end of Batch 0 (\textit{left}) is caused by the shear at the instrument carrier interface.} 
\end{figure}

In Fig.~\ref{fig:os11dh} we compare a dark hole image {in normalized intensity units} from the OS11 package to the corresponding dark hole generated with \tool~for the case of model uncertainty factors (MUFs) added. These images have been computed by coadding the images from OS11 batch 0, i.e. the first reference acquisition of the observing scenario. The normalized intensity is defined as the average flux in the dark hole divided by the peak flux of the off-axis PSF; it is an approximation of the raw contrast that ignores the effects of the FPM on the throughput. For \tool's simulations we use Phoenix stellar spectrum models \cite{Allard2003}. No detector noise was added. 
Fig.~\ref{fig:os11ni} shows a comparison of the normalized intensity in the dark hole between the OS11 product and the output of \tool.  Two OS11 batches are shown, a reference acquisition batch (batch 0) and the first roll of the science acquisition (batch 100).  

{The averaged normalized intensities in the dark hole are comparable, with a discrepancy of $\sim14$\%. For the purposes of this study, this difference is small, because the key objective is to ensure that changes resulting from an evolving wavefront are similar. Regarding the difference in speckle structure, small differences in the structure are to be expected; in practice, the Roman Coronagraph will have small differences in speckle structure every time the dark hole is re-computed. The main limitation in the speckle-limited regime, as will be seen in Sec.~\ref{sec:simus_description}, is the residuals of PSF-subtraction driven by differences in the wavefront errors: these differences are dominated by low-order aberration changes.}

\section{Effects of Jitter on Detection of Extended Exozodi Disks in Nearby Systems}\label{sec:tauceti_disk_simus}
Residual uncorrected jitter degrades contrast, particularly near the inner working angle. Time-variable jitter or jitter that varies between target and reference stars results in time- or target-variable speckle intensity and morphology, which in turn reduces the efficacy of PSF subtraction. This presents challenges for science cases involving weak signals, such as those the Roman Coronagraph intends to address. In this section we study the effect of jitter on the detection of exozodi disks in nearby systems, where the exozodi disk flux is primarily located outside the coronagraph IWA. Excessive jitter in a coronagraph instrument introduces two types of challenges. First, in the case where a weak disk signal is of similar intensity to that of the residual starlight after PSF subtraction, excessive jitter can cause a false negative. Second, the jitter-induced increased flux at the inner working angle raises the concern that it could be mistaken for disk flux, resulting in either a false positive disk detection or an erroneously-elevated disk flux measurement. In this section, we investigate the impact of various jitter scenarios on the detection and characterization of exozodiacal dust disks.

\subsection{Simulation Details}
\subsubsection{Case Study}\label{sec:taucet}
We choose \taucet~{as }a case study target to simulate the effect of jitter and drift on the detectability of exozodiacal light.
\taucet~is the closest Sun-like single-star with a comparable age to the Solar system. It hosts an extended outer debris disk 
\cite{Lawler2014,MacGregor2016}, and LBTI\cite{Ertel2020} observations placed a $\times$80 zodi upper limit near the habitable zone.
In the inner region a $\sim$1\% excess at 2.2 $\mu$m was detected with CHARA\cite{diFolco2007,Absil2013}; further observations with the Palomar Fiber Nuller\cite{Serabyn2019} were not able to detect it, which could be interpreted as evidence that the dust responsible for the excess emission is hot dust within 0.1 AU (Mennesson and Serabyn, priv. comm.).  
Given its proximity and similarity to the Sun, \taucet~is one of the priority targets for exoEarth candidate searches with HWO, as identified by ExEP \cite{Mamajek2024}. As such, placing constraints on the zodi cloud scattering properties in visible light near the habitable zone of \taucet~will inform the feasibility of an exoEarth search and characterization. Ref.~\citenum{Douglas2022}~studied a sample of nearby systems to assess the sensitivity of the Roman Coronagraph to reflected light from exozodi, concluding that \taucet~was one of the most favorable targets~as a result of its proximity. It is worth noting that we are not intending to self-consistently model the \taucet~system, with a disk that is consistent with the known planets. Rather, we are using the stellar parameters to create an illustrative model of exozodi sensitivity for a representative star. 

\subsubsection{Disk model}\label{sec:disk_model}
Exozodi disk models were constructed with the radiative transfer code \texttt{MCFOST} \cite{Pinte2006}{. The density of the disk is separated in two terms: the radial term, or surface density, and the vertical density. The} surface density distribution {follows a} functional form:
\begin{equation}
    \Sigma(r) \propto (r/r_{0})^{\alpha}
    \label{eq:surfacedensity}
\end{equation}
where $r$ is the radial coordinate, $r_{0}$ is a reference radius of 1.0 AU, $\alpha = -0.45$ is the radial surface density power law index, and vertical density distribution of the form
\begin{equation}
    \rho(r,z) \propto exp(\frac{-z^{2}}{2h(r)^{2}})
    \label{eq:verticaldensity}
\end{equation}
where $z$ is the vertical coordinate perpendicular to the disk midplane and $h(r)$ is the scale height of form
\begin{equation}
    h(r) = h_{0}(r/r_{0})^{\beta}
\end{equation}
where $h_{0} = 0.01$ is the reference scale height and $\beta = 1.25$ is the disk flaring index. The radial extents of the disk are truncated to a range of 0.1--10 au. The models are positioned with the disk midplanes horizontal, and the inclination is varied from 0 to 90$\degree$. For scattering properties, we apply the 3-component Henyey-Greenstein phase function fit to the Solar System zodiacal light described in Ref.~\citenum{Hong1985}{, known as the Hong phase function. The Hong phase function is better suited for this study than Mie-based scattering models; Mie-based scattering has been found to poorly approximate dust scattering properties (see e.g. Ref.~\citenum{Arriaga2020}). However, the Hong phase function is known to underestimate the scattering efficiency at small phase angles \cite{Hong1985}. This limitation is not critical in the context of coronagraphic imaging, since the corresponding scattering angles occur near the center of the FPM, where the stellar flux is heavily suppressed and the impact of such inaccuracies is minimized}. 
\texttt{MCFOST} was chosen to enable maximal flexibility in adjusting model parameters. While \texttt{ZODIPIC} \cite{Kuchner2012} has historically been used to model exozodiacal dust structure, surface density distribution and scattering formalisms are not adjustable. {The specific model parameters and the rationale for their selection are described in Ref.~\citenum{Lin2025}. The model agrees well with ZODIPIC at low inclinations, while moderate deviations appear at higher inclinations though not to a significant extent.} {The discrepancy at higher inclinations is expected since a more inclined disk will tend to underestimate the flux.}

We normalize the total flux of these models in terms of zodi level. We define a disk with a 1$\times$~zodi equivalent as in Ref.~\citenum{Stark2014}: {a surface brightness of 22 \vmagasec~around a Sun-like star, corresponding to the scattered-light brightness at the Earth-equivalent insolation.}
{Since we do not model the radiative transfer of dust grains with specific optical properties, the dust albedo is not accounted. We assume that the grain color is effectively flat across the bandpass since the models are evaluated only at a 10\% bandwidth. This assumption is further supported by simulations of cold debris disks presented in Ref.~\citenum{Anche2023}. The stellar spectrum is accounted for in the normalization of the flux; we use \texttt{pysynphot}.}

The disk model is then convolved with the Roman Coronagraph point response function (PRF). We use a 2D version of the method explained in Ref.~\citenum{Douglas2019}. The PRF is represented by a cube of PSFs spaced across the image plane field of view with two sampling patterns: a fine grid, separated {$\sim$1}
mas for the area within the IWA, and a more coarse elsewhere. The convolved disk image is added to the simulated speckle series. {The resolutions were chosen to by trial and error, minimizing the interpolation errors.}

In Sec.~\ref{sec:hot_dust}, we simulate the particular case of hot dust near the star, which is likely present around \taucet. For this case we use a simplified disk toy-model, consisting of a face-on ring with uniform surface brightness and sharp edges. 

\subsubsection{Speckle series simulation}\label{sec:simus_description}
To simulate realistic astrophysical scenes in which we would observe the system described above, we use the data from OS11 (see Sec.~\ref{sec:tool_description}). 
With these wavefront error data, we run simulations using \tool~for different amplitudes of jitter, i.e., changing the \texttt{jitter\_x}~and \texttt{jitter\_y} arrays containing the amplitude of residual jitter in the x and y directions. In order to keep these simulations both realistic and easy to compare between them, the different jitter scenarios are all taken from the OS11 jitter amplitude arrays and multiplied by different multiplicative factors. This scaling is referred to as the \textit{jitter parameter}, and it simply refers to the multiplicative factor by which the OS11 jitter amplitude array is multiplied. All other wavefront error and other parameters from OS11 are unchanged across simulations. In order to reduce computing time, we choose to simulate the first reference acquisition and first two telescope rolls science acquisitions (the two rolls are separated by 10$^\circ$). Since we aim at being in the speckle limited regime, rather than in the photon noise limited regime, these acquisitions are sufficient to provide enough SNR. In the OS11 package nomenclature these are batches 0, 100 and 101. These batches span almost 4 hours of observing time, which provides sufficient speckle diversity for our purposes.  

The speckle series resulting from the simulation described above is then used to create the final frames that include detector noise. This is done similarly to OS11: the noiseless speckle series is interpolated timestamps corresponding to individual exposures, and passed through the EMCCD simulator. We use \texttt{emccd\_detect}, version 2.4.0. Since the target star used in this work is different from OS11's, only the first batches of OS11 were utilized, and the exposure times are not optimized for our case study, we computed a different set of exposure times and number of frames. We use the Roman Coronagraph's exposure time calculator (ETC) tool, \texttt{EETC} {(\url{https://github.com/nasa-jpl/cgi-eetc})}. {The exposure time typically used for most simulations in this work is 30 sec per frame for \taucet, with $\sim$400 frames to achieve the desired SNR.} 
All frames at a given pointing (star and telescope roll angle) were coadded, resulting in three final frames: one reference star frame and one target star frame per roll.

\begin{figure}
   \begin{center}
   \begin{tabular}{c} 
   \includegraphics[height=8.0cm,trim={5cm 4cm 5cm 4cm}]{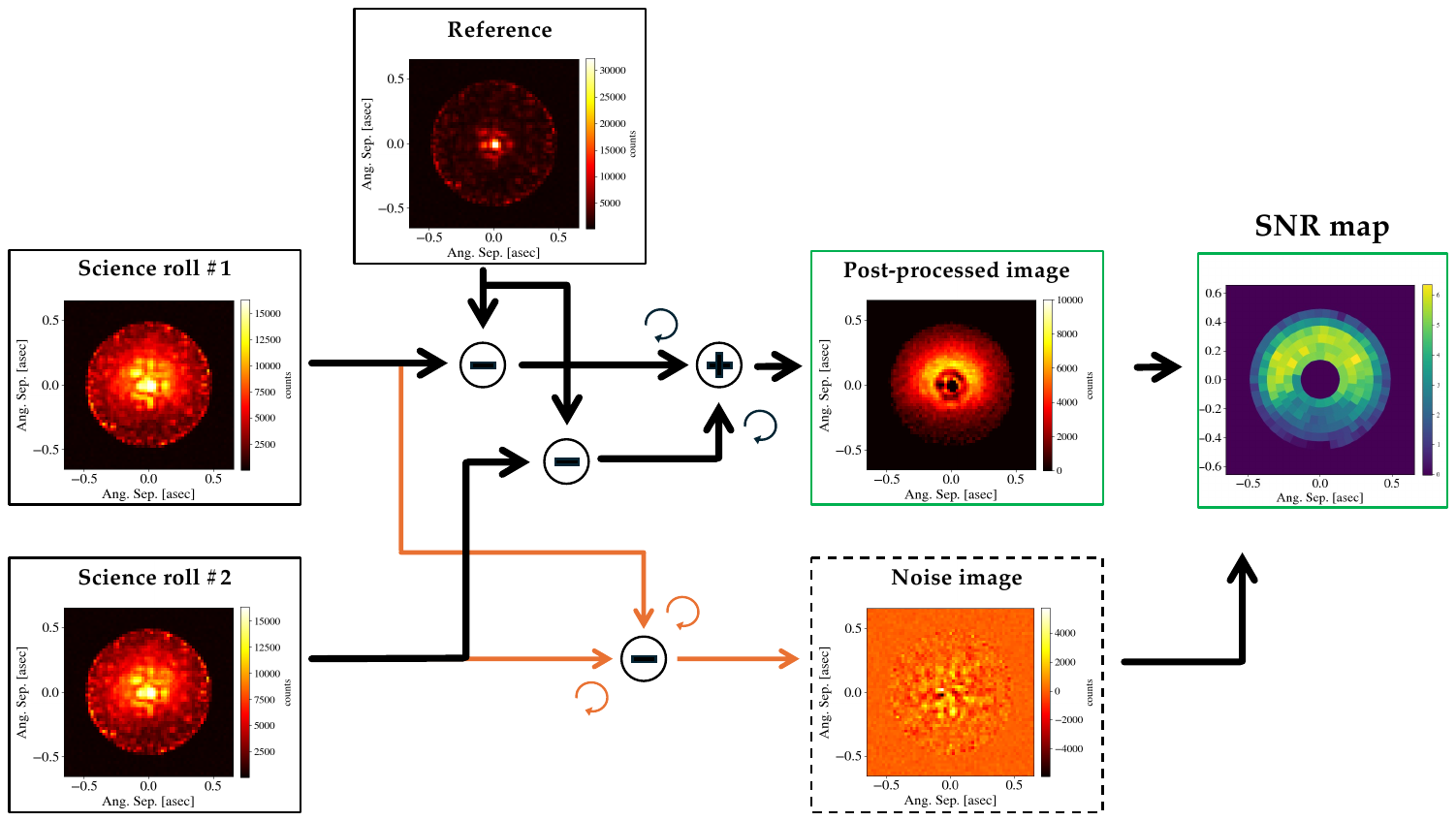}
   \end{tabular}
   \end{center}
   \caption{Flowchart illustrating the RDI observation strategy and post-processing steps that are taken to compute a final, post-processed image and an SNR map.
   \label{fig:rdi_snr}
   }  
\end{figure}

The reference acquisition is obtained to perform reference differential imaging (RDI), where the reference is used as a PSF reference to subtract the diffraction speckles. {The reference star for all simulations is $\zeta$ Puppis, an O4I star with a stellar diameter of 0.9 mas, all of which is accounted in the simulations. We assume no dust is scattered near the reference star. }
The two telescope rolls simulated in OS11 were designed for angular differential imaging (ADI), where the coadded images obtained in each of the 2 roll states are generally subtracted from one another to reduce the speckle noise. However, in the case of disk science, ADI risks subtracting disk signal when the roll subtraction is performed. Therefore, the two telescope rolls are only de-rotated and coadded without performing ADI. Before coadding, the reference image is subtracted from the roll images independently. Fig.~\ref{fig:rdi_snr} shows a schematic of the post-processing steps. Ideally, if the telescope was perfectly stable and residual jitter amplitude was time-constant and independent of star brightness, the speckles from the reference image and those from the science rolls {would only differ due to stellar characteristics}. However, the level of telescope and instrument variability captured in the OS11 wavefront error time series is sufficient to introduce noticeable changes in speckle structure between reference and science observations. As a result, residual speckle noise from imperfect subtraction remains the primary limitation in the final PSF-subtracted and coadded images.

In order to compare results from different scenarios, we use a signal-to-noise (SNR) metric. An SNR map is computed by measuring the mean flux in small regions, or sections, in the final, post-processed image. The sections are separated in annuli centered on the star. Each annulus is divided in sections, the inner side of which is no smaller than the FWHM of the off-axis PSF; Fig.~\ref{fig:rdi_snr} shows an example of an SNR map. The signal is the average flux within a section, which includes both disk flux and residual speckle flux. Regarding the noise contribution to the SNR, we employ an approach similar to one demonstrated in Refs.~\citenum{Hom2024} and \citenum{Mazoyer2020}. A noise image is computed by self-subtracting the extended emission signal. 
This is done by first subtracting the reference image from the science rolls independently. Then, instead of coadding both PSF-subtracted, de-rotated rolls as is done to compute the final science image, one PSF-subtracted, de-rotated roll is subtracted from the other. This removes {the disk signal}.
This noise image is then used to compute the noise value by calculating the standard deviation {with the resolution elements available} at the same separation as the section in which the flux is extracted. {This assumes that the noise is Gaussian, which is an approximation made to simplify the calculations; the discrepancy is larger in the inner region where the speckle noise dominates \cite{Pairet2019}. In the inner region we only dispose of 16 independent resolution elements to compute the standard deviation, which would require a small correction to account for small sample statistics\cite{Mawet2014}, this is not accounted for in our evaluation.} {In our simulations the speckle noise typically dominated within $\sim$200 mas.}
The final SNR number reported for an observation is the maximum value in the SNR map{; a mean over the area covered by the disk would require prior knowledge on the disk, and would be sensitive to the accuracy of that knowledge. To avoid this dependence, we report the maximum value, thereby leveling the comparison or mimicking a blind search. In practice, observers will likely have some prior information about the disk, enabling a more precise assessment of the detection confidence. Our} SNR metric represents a straightforward way of comparing different scenarios that we simulate. 
Even though the noise image is not an exact representation of the underlying noise present in the final image, we believe it is preferable to have a common methodology for calculating SNR across all scenarios. We find that the SNR values of 5 and above provide a reasonable threshold of detection when inspecting the model images and the final images. 

\subsection{Exozodi Simulation Results}

\subsubsection{Detectability as a functions of inclination and zodi level}
\begin{figure}
   \begin{center}
   \begin{tabular}{c} 
   \includegraphics[height=19.0cm]{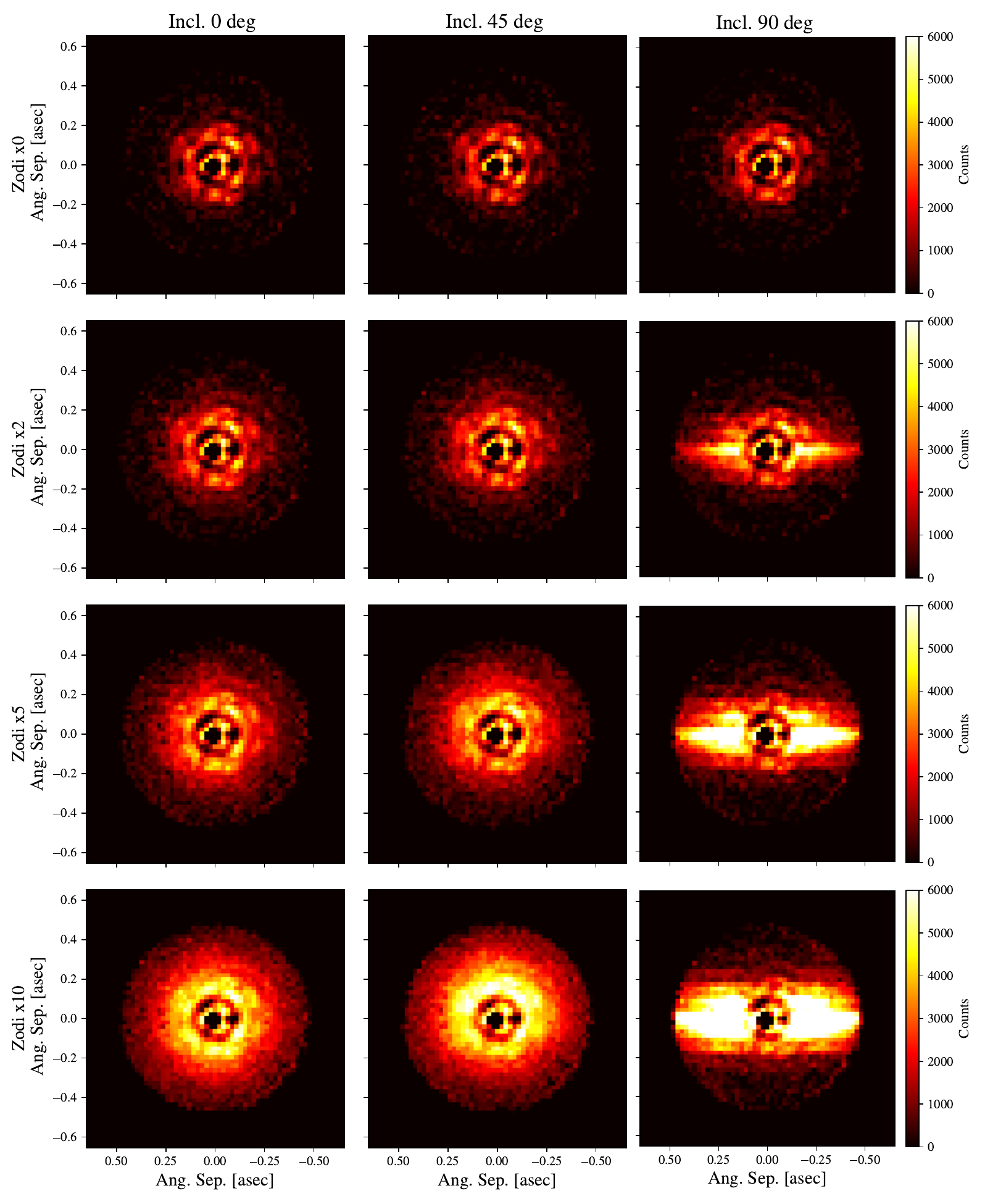}
   \end{tabular}
   \end{center}
   \caption{
   \label{fig:inc_vs_zodi}
   Final images of simulated observations of \taucet's~disk for different levels of zodi and inclinatons of the disk. The disk model, described in Sec.~\ref{sec:disk_model} is convolved with the point response function (PRF) and added to a speckle series, generated as described in Sec.~\ref{sec:simus_description}. The simulations include a reference observation for PSF subtraction and two rolls, these images show the PSF subtracted product with the two rolls combined. These simulations include detector noise. As expected, at lower inclinations the disk is harder to detect and the signal gets mixed with the residual speckles of comparable brightness. Note that the top row provides a visualization of the images obtained with no exozodi present, i.e. post-calibration speckle residuals only.}  
\end{figure}

We simulate the case of an exozodi with \taucet's stellar parameters for different inclinations and different levels of zodi to assess what sets of parameters make it detectable under our assumptions. Fig.~\ref{fig:inc_vs_zodi} shows images from simulated scenarios to illustrate the interplay between amount of dust, quantified here as zodi level, and inclination. The higher the inclination, the more it enhances the forward scattering, which makes the detection of the disk easier. It is worth noting that this work is focused on disk detection as opposed to mitigation of disk contamination to search for planets; edge-on or clumpy disks may resemble and be confused with point-sources. 

\begin{figure}[t!]
   \begin{center}
   \begin{tabular}{c} 
   \includegraphics[height=7.25cm,trim={2cm 0cm 2cm 0cm}]{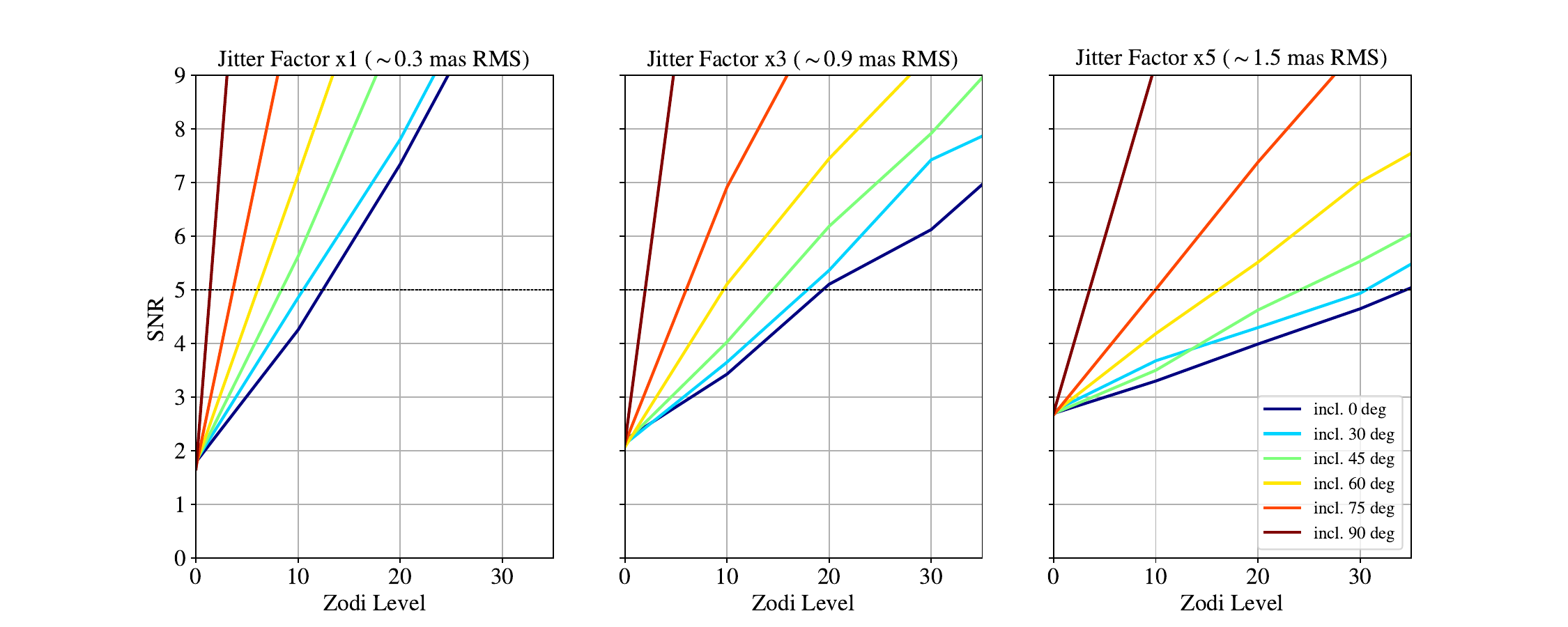}
   \end{tabular}
   \end{center}
   \caption{
   \label{fig:snr_vs_zodi}
    SNR as a function of zodi level, or brightness of the disk, for different disk inclinations and target star jitter levels. RDI reference star has the same nominal jitter time series in all cases, with a jitter amplitude of $\sim$0.3 mas RMS. Higher jitter hinders the detectability of the disk by increasing the intensity of the final residual speckles. This effect is more pronounced for lower inclination disks.
   } 
\end{figure}

In Fig.~\ref{fig:snr_vs_zodi} we show SNR for different inclinations (0\degree - 90\degree), zodi levels (1 - 35~zodi), and target star average jitter amplitudes (0.3~mas - 1.5~mas RMS). The reference star average jitter remained fixed at 0.3~mas in all cases. The SNR reported in this figure is the largest per-section SNR number in a final, post-processed image. The definition of this SNR metric is explained in Sec.~\ref{sec:simus_description}. As expected, the inclination of the disk has great impact on the detectability of the disk. For our case study, \taucet, and the disk model assumptions (see~Sec.~\ref{sec:disk_model}), the zodi level needed for detection (SNR$>$5) is as low as $\sim$1~zodi for the edge-on case in the nominal jitter case, but for a face-on inclination, the disk must be brighter than $\sim$12~zodi. {At zero zodi (no disk present), the SNR is set by residual starlight; increasing jitter raises the apparent SNR in the no-disk case.}

\subsubsection{Detectability with increased levels of jitter}
\begin{figure}[t!]
   \begin{center}
   \begin{tabular}{c} 
   \includegraphics[height=8.0cm]{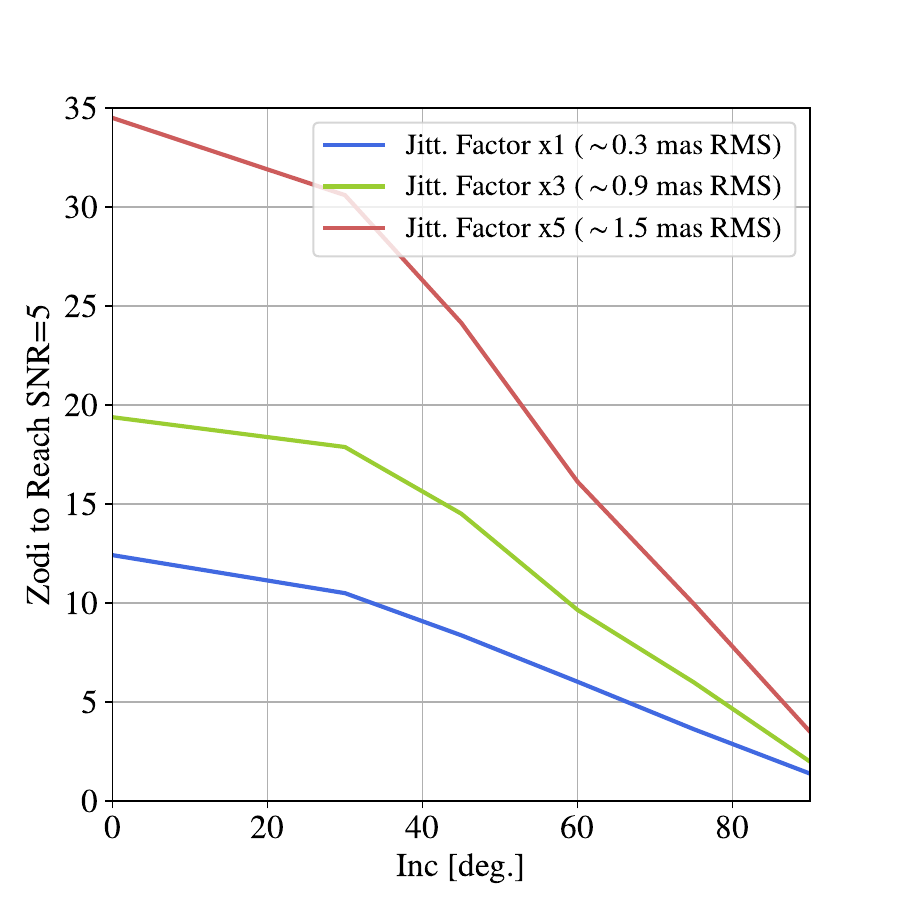}
   \end{tabular}
   \end{center}
   \caption{
   \label{fig:zodi2reachSNR5}
    Zodi level of a disk needed to reach SNR=5 as a function of disk inclination, for different levels of telescope jitter on the target star. For lower inclination disks, the zodi level, or brightness of the disk needed is significantly higher, and is more affected by the presence of jitter.
   } 
\end{figure}

The effect of increasing jitter is the increase of the noise after post-processing, which hinders the detectability of a disk. In the three panels of Fig.~\ref{fig:snr_vs_zodi}, we show three cases of target star jitter levels: the zodi level needed for a detection increases with the jitter amplitude. Another way of plotting these results is by plotting the zodi level needed to reach SNR=5 for different inclinations; we show this in Fig.~\ref{fig:zodi2reachSNR5}. As expected, the detectability is more affected by the jitter level for face-on disks. For instance, in the case of the 30\degree~disk, a zodi level of $\sim$7 is needed for detection in the nominal scenario, factor $\times$1, but for the case of large jitter, factor $\times$5, a zodi level of $\sim$13 is required for detection. For high inclination disks, from 75\degree~to 90\degree, the difference is smaller.

\begin{figure}
   \begin{center}
   \begin{tabular}{c} 
   \includegraphics[height=7.5cm]{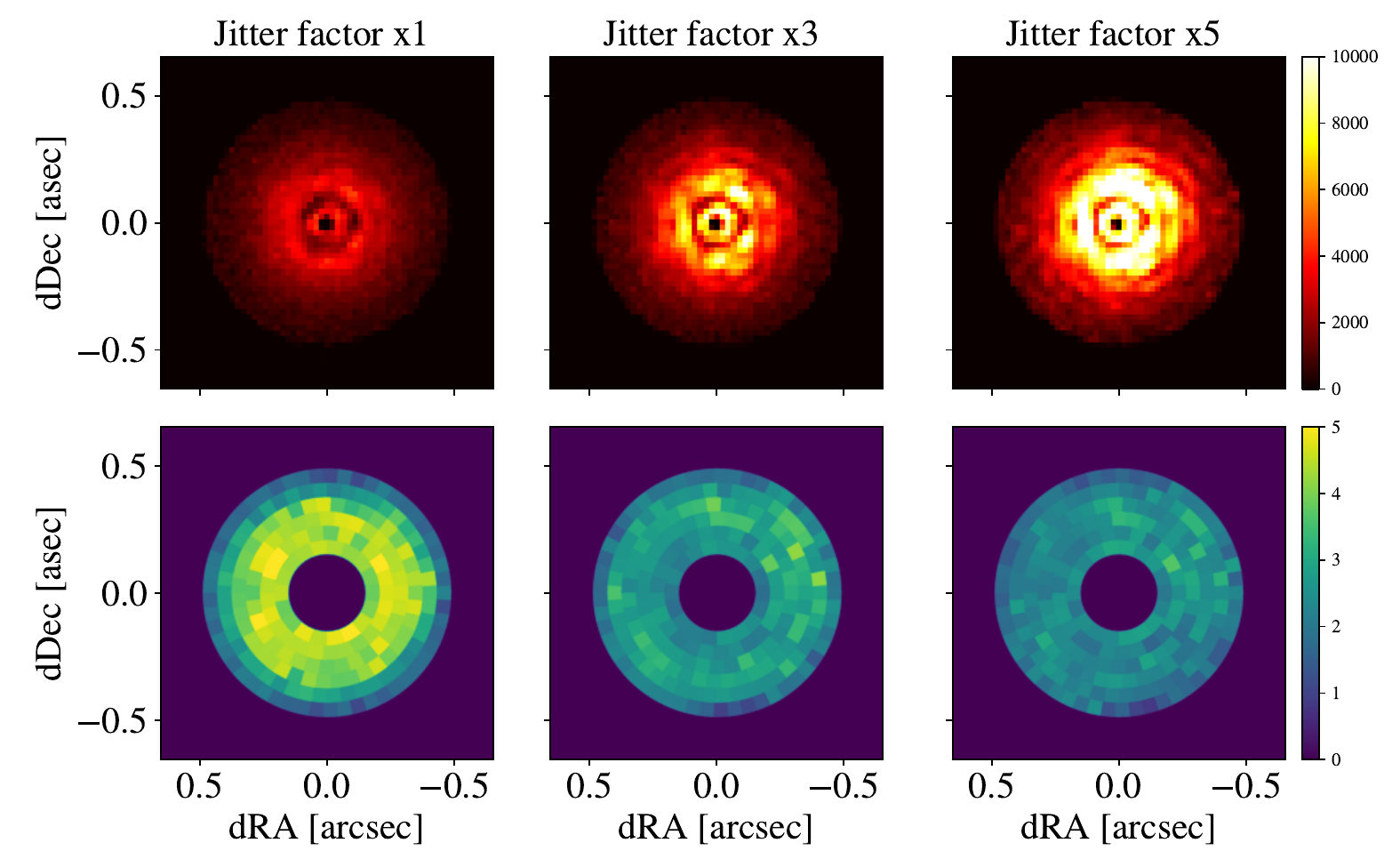}
   \end{tabular}
   \end{center}
   \caption{
   \label{fig:images_and_snr}
   \textit{Top:} simulated observations of the \taucet~exozodi disk (see details of these simulations in Sec.\ref{sec:simus_description}) for a disk of 
  22x zodis, for different levels of jitter. \textit{Bottom:} map of the SNR values corresponding to the above image. The increased level of jitter introduces residual speckles that hinder the detection of the disk. This noise is largely captured by our SNR metric, which avoids false positive signals for this level of jitter. For details on the SNR metric used here, see Sec.~\ref{sec:simus_description}. The zodi level was chosen to give an SNR value of $\sim$5 for the nominal case of jitter factor $\times$1.
   } 
\end{figure}

In Fig.~\ref{fig:images_and_snr} we show the case of a 22~zodi exozodi disk that is detectable at the nominal level of target star jitter (0.3~mas RMS), but its signal becomes overwhelmed by the speckle residual when increasing the level of jitter. The zodi level was chosen to give an SNR value of $\sim$5 for the nominal jitter value. {The non-uniformity in the SNR map is due residual starlight after post-processing.}

\subsubsection{Possible confusion of jitter speckle structure with disk}
\begin{figure}[t!]
   \begin{center}
   \begin{tabular}{c} 
   \includegraphics[height=7.5cm,trim={2cm 0cm 2cm 0cm}]{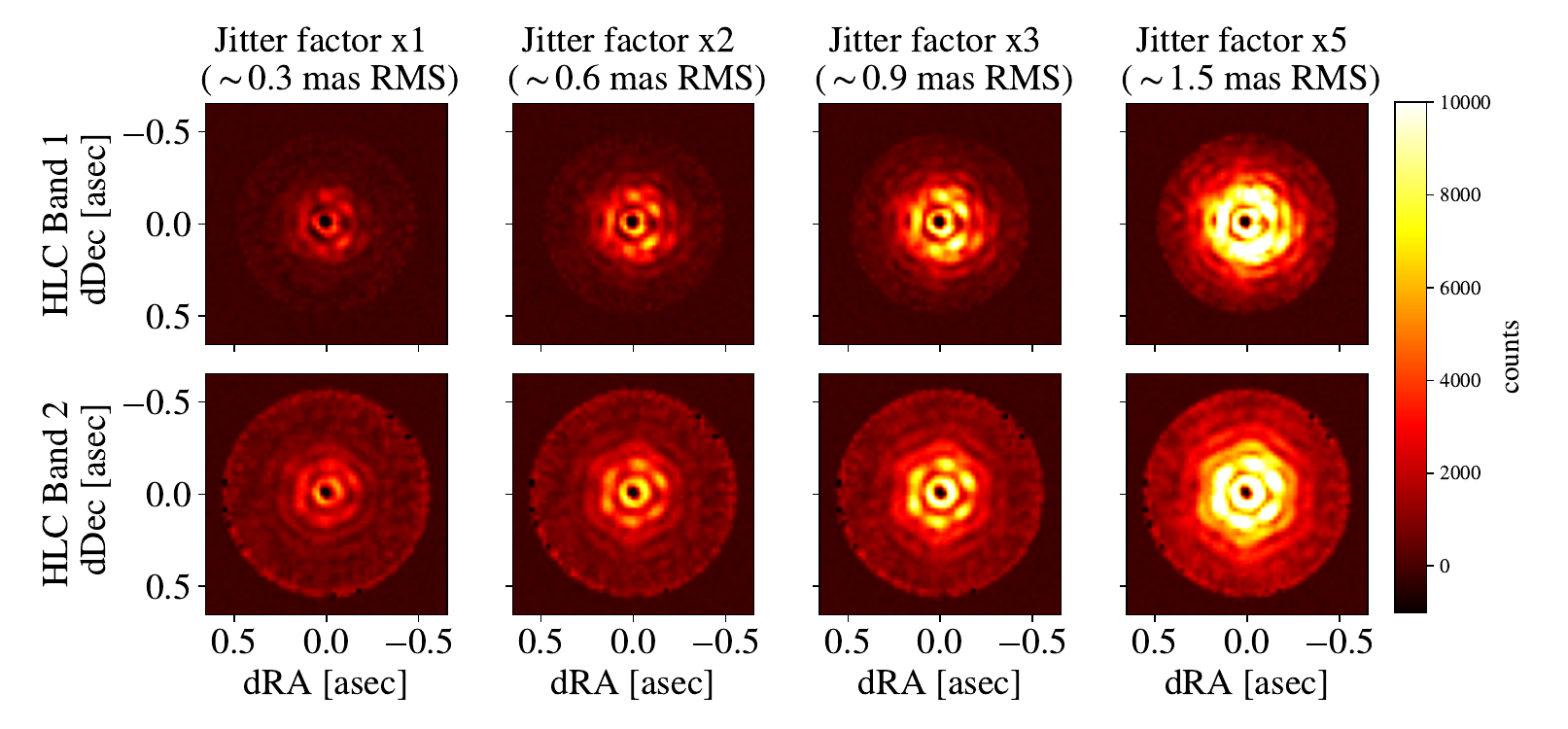}
   \end{tabular}
   \end{center}
   \caption{Post-processed images for increasing levels of target star jitter for HLC Band 1 (\textit{top}) and HLC Band 2 (\textit{bottom}). These images are the result of the observation scenario and post-processing  
    described in Sec.~\ref{sec:simus_description}. These images illustrate that the jitter signature has a distinct, non-azimuthally symmetric structure. Furthermore, the speckle structure is not consistent between Band 1 and Band 2, which would not be the case for disk structure. 
   \label{fig:dhs_no_disk}
   } 
\end{figure}

The other key concern when addressing jitter is whether one is able to discern the difference between an astrophysical signal and residual starlight. 
In Fig.~\ref{fig:dhs_no_disk} we show a set of post-processed images resulting from the observation scenario described in Sec.~\ref{sec:simus_description} with increasing jitter and zero disk signal.
We find the jitter induced speckle structure is easily identifiable.
This is due to the concentration of energy close to the center of the FPM, which results in a few bright speckles near the IWA, as opposed to a disk with a more uniform and extended distribution of energy.   Fig.~\ref{fig:images_and_snr} illustrates a similar point: the increased residual jitter is unable to trigger a high SNR when looking for a flat disk that extends beyond the IWA.
In Sec.~\ref{sec:hot_dust}, we explore the worst case scenario with a simplified toy model of a very small disk that would almost perfectly mimic jitter.

A solution to mitigating confusion is to obtain color diversity with additional imaging bandpasses.  HLC Band 2 ($\lambda_0 = 660$~nm) is the natural complementary imaging bandpass\cite{Riggs2021} {since} it has also has an IWA of 3 $\lambda/D$. It has its own FPM {but it is not a supported mode} and needs a different DM solution. In the bottom row of Fig.~\ref{fig:dhs_no_disk} we show the corresponding images {with HLC Band 2 to compare them to the above images of the HLC Band 1 mode}. At all jitter levels, the speckle morphology is distinct between HLC Band 1 and HLC Band 2, whereas the morphology of any true clumpy disk structure would be {nearly identical} at both wavelengths. {Differences in bandpass, DM solution, and FPM inevitably produce a mismatch in speckle structure between the two bands. This discrepancy can be exploited, for example, by jointly fitting models to both bands with priors on the expected chromatic behavior of the disk, thereby improving discrimination between stellar speckles and genuine disk signal.}

\subsection{Hot dust toy model}\label{sec:hot_dust}
\begin{figure}
   \begin{center}
   \begin{tabular}{c} 
   \includegraphics[height=9.0cm]{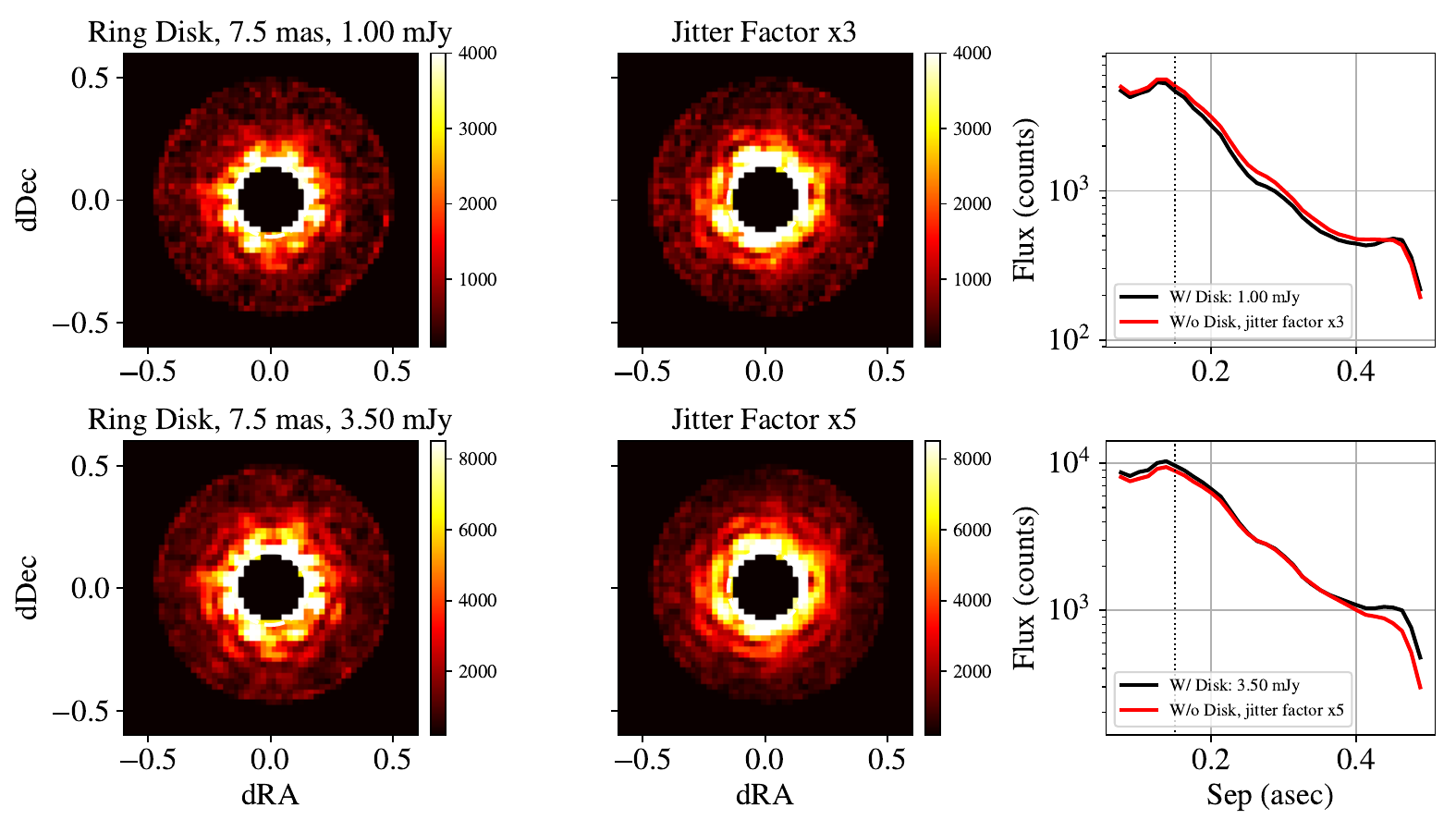}
   \end{tabular}
   \end{center}
   \caption{
   \label{fig:toyModel_confusion}
    Comparison of post-processed HLC Band 1 images of two hot exozodi toy models with nominal jitter (\textit{left~column}) versus two higher jitter cases without disks (\textit{middle~column}). The plot on the \textit{right~column} compares the retrieved flux curve in both cases; the two curves match each other well, indicating the possible confusion between jitter and hot dust. Indeed, the disk diameter (15~mas) is $<\lambda/D$ and is not distinguishable from jitter. This suggests that hot zodi should be considered when choosing PSF reference stars for Roman Coronagraph. {For reference, the flux curve corresponding to the nominal case of jitter and no disk, not plotted here, would be $\sim$3 times lower flux than the bottom curves. }
    }
\end{figure}

In this section we simulate the particular case in which the disk is $<\lambda/D$ from the star. For instance, in the case of \taucet, hot dust is likely present within 0.1 AU (see Sec.~\ref{sec:taucet}). This is a particularly interesting case because that the projected size on the focal plane is of comparable size to the jitter amplitude. Indeed, the jitter signature in the form of residual speckles in the image plane is effectively equivalent, with sufficient temporal sampling, to the image of a small, {face-on} extended source. To illustrate this, we simulate observations of a simplified disk toy-model which brightness and appearance in the image plane mimic that of high jitter structure. In Fig.~\ref{fig:toyModel_confusion} we show two simulations comparing the case of the toy-model disk, with a nominal amount of jitter, versus a case of no disk, but with an increased level of residual jitter. 
The simulations of these observation follow the procedure described in Sec.~\ref{sec:simus_description}: two science rolls and a reference observation for PSF subtraction. 

The jitter signature mimics the small disk because both image sets are computed similarly. Effectively, the jitter speckles are a convolution of a 2-D Gaussian function with the Roman Coronagraph PRF, and the disk in this instance is the convolution of a top-hat function with the same PRF. To match the intensities, the increased jitter scenario with a jitter factor of $\times$3 requires a 1 mJy total flux disk, whereas a jitter factor of $\times$5 requires a 3.5 mJy disk. These flux numbers correspond to a flux ratio of $\sim10^{-6}$ with respect to the flux from the star. {This result applies to a face-on disk; if the disk is inclined the convolved structure will show the azimuthal variation of the disk, as shown in Fig.~\ref{fig:inc_vs_zodi}.}

\section{Measured Ring Separation and Flux Bias for Disks Inside the IWA}\label{sec:toymodels_degeneracy}
\begin{figure}
   \begin{center}
   \begin{tabular}{c} 
   \includegraphics[height=7.5cm,trim={1.6cm 5cm 1.2cm 5cm},clip]{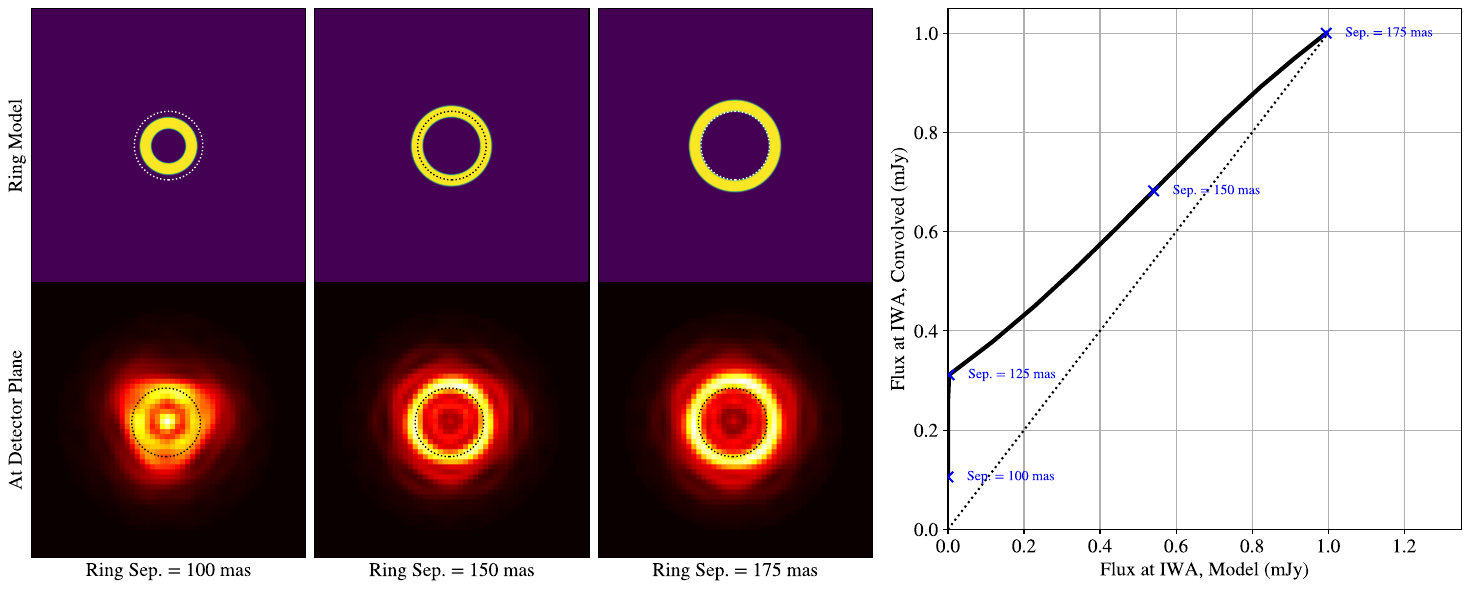}
   \end{tabular}
   \end{center}
   \caption{
   \label{fig:toyModel_images}
    \textit{Left-top:} Ring disk toy models. The dotted circle denotes the IWA. \textit{Left-bottom:} Convolved image of the above model with the HLC Band 1 point response function. The intensity is biased outward, toward the edge of the FPM, for ring separations less than the IWA. This bias is inherent to this type of coronagraph FPM. \textit{Right:} Flux recovered from the convolved image versus the flux from model for different ring separations. The flux is computed in the inner-most region: one resolution element the inner edge of which is the IWA. When the model intensity is zero, i.e. when the ring model is fully outside the flux-extraction region, some residual flux is recovered in the convolved imaged. The dashed line represents if there was perfect knowledge of the model. At 175 mas, when the model ring fully falls outside the FPM outer edge, the flux correction matches the correction needed to recover the flux perfectly.} 
\end{figure}

In Ref.~\citenum{Krist2023}, the effect of PSF truncation near the IWA is shown to produce an offset between the measured position of a point source and its actual position. This is caused by the sharp edge of the focal plane mask: the centroid of the PSF is effectively shifted by the partial blocking of the PSF at the FPM; the Lyot stop re-images the PSF with a slight offset. We simulate the case of a thin dust ring (hot or warm exozodi toy model) within or close to the edge of the HLC Band 1 FPM. Fig.~\ref{fig:toyModel_images} shows the ring models and the convolution with the instrument's PRF. The ring model has a width of 50 mas and its separation is defined as the separation of its center.
Unlike previous figures shown in this paper, this figure shows only the ring model, without the host star; it is not a simulated science image with noise and jitter. 
The curve shown in this figure represents the flux recovered {via aperture photometry} near the IWA from a convolved image versus the true flux from the model. The flux recovered from the convolved images is corrected to match the true flux accounting for throughput losses. For rings of small separations, the flux mostly falls under the FPM {and is \textit{displaced} from its true location outwards in the camera plane. This results in a degeneracy that can complicate forward modeling efforts and deconvolution; prior knowledge on the disk brightness and structure can mitigate this effect. }
 
\begin{figure}
   \begin{center}
   \begin{tabular}{c} 
   \includegraphics[height=8.0cm]{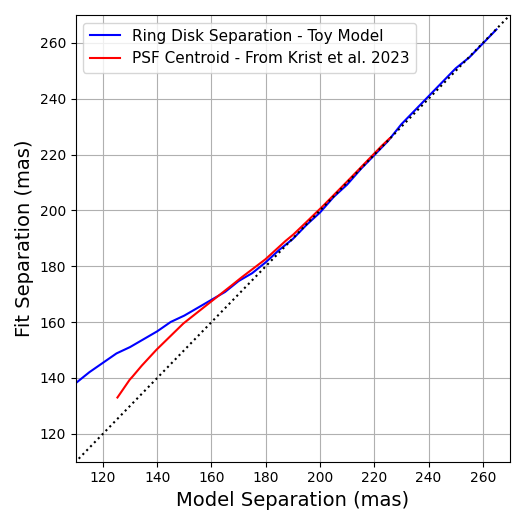}
   \end{tabular}
   \end{center}
   \caption{
   \label{fig:toyModel_curves}
    Ring disk toy model measured separation (\textit{blue}) after convolution with the HLC band 1 point response function (as shown in Fig.~\ref{fig:toyModel_images}) vs. its true (``model'') separation. The point source (``PSF'') centroid measurement reported in Ref.~\citenum{Krist2023} is also plotted (\textit{red}). The difference in behavior between the point source and ring cases can be attributed to a compounding-like effect of dealing with a ring as opposed to a point-like source.
   }
\end{figure}

The offset of the measured separation with respect to the true model separation of the ring is plotted in Fig.~\ref{fig:toyModel_curves}. The offset computed in Ref.~\citenum{Krist2023} for the point-source PSF is also displayed. The offset for the case of the ring seems to be worse for small separations; this can be attributed a \textit{compounding} effect of the response function. 

This bias would not only affect the disk separation but the inclination as well. For instance, a flat, face-on disk that extends within the FPM could be attributed a higher inclination when the light reflected closer to the star is pushed to the edge of the FPM and makes it look brighter and more inclined. This toy model underscores the criticality of forward modeling to the interpretation of disk observations at small separations; more detailed exploration of disk parameter recovery is left to future work. 

This negative effect is solely due to the nature of the FPM. Indeed, a hard edge FPM will always introduce this degeneracy, not only for point source separation but for disk inclinations. Coronagraph architectures that do not make use of a hard edge FPM, such as the vortex coronagraph~\cite{Mawet2005}, are largely unaffected by this weakness. When considering reflected light, improved constraints on the inclination of a disk may prove invaluable. For instance, when considering a follow-up on a newly detected point source, a disk detection with a well constrained inclination may provide free constraints to the planet inclination. With this information, the scheduling of follow-up can be better optimized.  

\section{Conclusions}
In this article, we presented a new tool, \tool, to simulate observations with the Roman Coronagraph and allow for convenient propagation of astrophysical scenes with an evolving amount of arbitrary wavefront errors, including residual pointing jitter and drift. This package wraps around \cgisim, and was validated to output similar results to OS11 with similar inputs. 

\tool~was used to simulate observations of hot and temperate exozodi disks, with different assumptions for the disk inclination and instrumental jitter. The main conclusions are summarized as follows:

\begin{itemize}
    \item Simulations with \tool~of observations of a smooth exozodi disk in the \taucet~system indicate that the Roman Coronagraph would be sensitive to $\sim$12$\times$ zodi levels, assuming a face-on disk and with conservative (MUF'd) instrument wavefront control performance. In the case of edge-on inclination, a zodi level of $\sim$1$\times$ is needed for detection.
    \item For an RDI scenario with increased jitter on the target star compared to its PSF reference star, the additional speckle noise hinders the detectability of temperate exozodi disks. This effect is exacerbated for lower inclination disks. 
    \item Although confusion between speckles originated by residual jitter and a face-on extended disk is possible, we find that it is unlikely for disks, such as temperate exozodi, extending well beyond the coronagraph IWA. The confusion can be further mitigated by two-color imaging. 
    \item Confusion between jitter-induced speckle residuals and hot exozodi disks, whose projected separations can be $<\lambda/D$, is possible for sufficiently faint disks, and is worthy of future study.
    \item Flux originating from dust located within IWA can ``leak'' outside the IWA due to the hard-edge effect of the coronagraph FPM. This flux/position degeneracy should be accounted for when interpreting disk images. 
\end{itemize}

\section*{Code and Data Availability}
The code presented in this paper can be found on Github: \url{https://github.com/jorgellop/corosims}. The data utilized to run some of the simulations can be freely accessed through Zenodo: \url{https://doi.org/10.5281/zenodo.15586669}.

\section*{Disclosures}
The authors declare there are no financial interests, commercial affiliations, or other potential conflicts of interest that have influenced the objectivity of this research or the writing of this paper.
\section*{Acknowledgements}
Part of this work was carried out at the Jet Propulsion Laboratory, California Institute of Technology, under a contract with the National Aeronautics and Space Administration (80NM0018D0004) and funded through the internal Research and Technology Development program. This research has made use of the SIMBAD database, operated at CDS, Strasbourg, France. AI tools were used for language and grammar clean-up. 
\copyright 2025. All rights reserved.

\bibliography{references.bib}{}
\bibliographystyle{spiejour}

\listoffigures
\listoftables


\end{spacing}
\end{document}